\documentclass[11pt,a4paper]{article}

\usepackage[utf8]{inputenc}
\usepackage[T1]{fontenc}
\usepackage{amsmath,amssymb,amsthm}
\usepackage{bm}
\usepackage{hyperref}
\usepackage{geometry}
\usepackage{xcolor}
\usepackage{caption}
\usepackage{subcaption}
\usepackage[normalem]{ulem}
\usepackage{graphicx}
\usepackage{float}
\usepackage[
backend=biber,
style=nature,
sorting=none,
sortcites=true,
maxnames=5
]{biblatex}
\bibliography{References.bib}

\definecolor{darkgreen}{rgb}{0.0, 0.5, 0.0}

\begin{document}

{\renewcommand{\thefootnote}{\fnsymbol{footnote}}
        
\begin{center}
{\LARGE Hidden quantum-informatic symmetries \\ of quasi-de Sitter backgrounds} 
\vspace{1.5em}

Suddhasattwa Brahma$^{1,2}$\footnote{e-mail address: {\tt suddhasattwa.brahma@gmail.com}}, 
Jaime Calder\'on-Figueroa$^{3}$\footnote{e-mail address: {\tt jrc43@sussex.ac.uk}},
Xiancong Luo$^{2}$\footnote{e-mail address: {\tt X.Luo-35@sms.ed.ac.uk}}, and
Vincent Vennin$^{4}$\footnote{e-mail address: {\tt vincent.vennin@phys.ens.fr}}
\\
\vspace{1.5em}

$^1$Physics and Applied Mathematics Unit, Indian Statistical Institute,\\ 203 B.T. Road, Kolkata 700108, India\\[2mm]

$^{2}$Higgs Centre for Theoretical Physics, School of Physics and Astronomy,\\ University of Edinburgh, Edinburgh, EH9 3FD, UK\\[2mm]

$^3$Astronomy Centre, University of Sussex, Falmer, Brighton, BN1 9QH, UK\\[2mm]

$^4$Laboratoire de Physique de l’Ecole Normale Sup\'erieure, ENS, CNRS, Universit\'e PSL, Sorbonne Universit\'e, Universit\'e Paris Cit\'e, F-75005 Paris, France

\vspace{1.5em}
\end{center}
}

\setcounter{footnote}{0}

	\begin{abstract}
		We investigate how degeneracies in quasi-de Sitter backgrounds, in the sense of Wands' duality, are reflected in real-space quantum correlations of primordial perturbations.	Using the continuous-variable Gaussian formalism for coarse-grained scalar fluctuations, we construct the covariance matrix of a pair of spatially localized modes in inflationary spacetime, and extract the symplectic invariants of the system.	For a generic Wands-dual pair of backgrounds, we find that while the individual entries of the covariance matrix are highly background-dependent, the symplectic eigenvalues -- and hence the entanglement entropy, mutual information, quantum discord and log-negativity -- all coincide for the two dual realizations. Our results unveil a new ``quantum-informatic symmetry'' of the de Sitter vacuum, according to which local linear entanglement witnesses constructed from coarse-grained fields cannot distinguish between Wands-dual inflationary histories, even though their background trajectories differ. We show that the special nature of the Wands-duality symmetry (of being local, scale-independent canonical transformations) is at the heart of this duality.
       

    \end{abstract}
	

	\section{Introduction}
	\label{sec:intro}
	
	Inflation offers a controlled setting to study quantum fields in an accelerating spacetime, with primordial inhomogeneities originating from vacuum fluctuations that are stretched to cosmological scales \cite{Mukhanov:1981xt, Mukhanov:1982nu}. A lingering question that remains unanswered is whether there is an observational probe that can identify the quantum origin of these fluctuations, thereby supporting the standard inflationary paradigm \cite{Campo:2005sv,Maldacena:2015bha, Martin:2015qta,Martin:2016tbd,Martin:2017zxs,Green:2020whw, Espinosa-Portales:2022yok, Sou:2024tjv}.  This is precisely where recent relativistic quantum information techniques provide a natural language to quantify how much of the ``quantumness'' of these fluctuations survives until late times, and how it might get encoded in observable correlations \cite{Kanno:2016gas, Hollowood:2017bil, Gomez:2020xdb, Martin:2021znx, Brahma:2021mng, Gomez:2021hbb, Gomez:2021yhd, Colas:2022hlq, Colas:2022kfu, Brahma:2022yxu, Brahma:2023hki, Chen:2024dcc, Brahma:2024ycc, Brahma:2024yor, Burgess:2024eng, Colas:2024xjy, Cielo:2025ibc, Lopez:2025arw, Piotrak:2025zhy}.
    
   
    A particularly useful framework relies on extracting two local quantum modes from a quantum field in a quasi-de Sitter background, and studying quantum correlations between them~\cite{Martin:2021xml, Martin:2021qkg}.
    For a free scalar field in a homogeneous and isotropic background, the vacuum state is Gaussian hence the two local modes are described by continuous-variable Gaussian states, whose
    statistical properties are encoded in a covariance matrix constructed from two-point functions in real space.	
    That covariance matrix fully characterises correlations between the two modes, and allows one to compute \textit{e.g.,}~
    entanglement entropy, mutual information, quantum discord, or log negativity between them \cite{Martin:2021qkg, Weedbrook:2011wxo, Wang:2007, Agullo:2024cln}.
	
	Most existing works along these lines have focused on standard slow-roll inflationary backgrounds and have addressed questions such as:
    How much entanglement is present between two localized regions in de Sitter space? Or, how the late-time entanglement structure of localized observables in an inflating universe compares with that in Minkowski space \cite{Martin:2021qkg, Martin:2021xml, Agullo:2024cln, Ribes-Metidieri:2025nfw}? In particular, by comparing results in flat \textit{vs} de Sitter space, it has been argued that inflation does not create entanglement between local modes \cite{Agullo:2024cln}. This confirmed previous analysis showing that traces for any such quantum correlation would be negligible in the resulting CMB \cite{Martin:2021qkg}. The inability of real-space entanglement entropy to distinguish between Wands-dual inflationary and matter-contracting phases, unless the measure itself was adapted to include characteristics of the background evolution, was previously noted in \cite{Chandran:2024utf}\footnote{On a related note, introducing Planck scale physics to distinguish between Wands duals have also been advocated \cite{Chandran:2025azu} although whether such trans-Planckian physics should affect any observables remains a topic of contention \cite{Burgess:2020nec}.}.

	In this paper, we pursue a complementary direction.	We ask how sensitive are such real-space quantum-informatic quantities to background degeneracies, in the precise sense of Wands' duality \cite{Wands:1998yp}. Wands' duality relates distinct inflationary histories that lead to identical linear dynamics for the Mukhanov-Sasaki variable. Different realizations of the background with different Hubble-flow parameters can generate the same effective mass term in the Mukhanov-Sasaki equation and, consequently, the same power spectrum for the canonical configuration field variable \cite{Wands:1998yp}. However, from the perspective of  cosmological observables, these dual backgrounds are certainly distinguishable since they lead to different curvature perturbations. A prototypical example of this duality is that of slow-roll and ultra-slow roll inflation, which forms a Wands dual, with identical Mukhanov-Sasaki configuration field but different background trajectories
    (see Refs.~\cite{Brahma:2024ycc, Cielo:2025ibc, Ireland:2026txt} for analyses on quantum aspects of inflationary perturbations in non--slow-roll backgrounds). Also note that  Wands-dual backgrounds differ in their momenta conjugate to the canonical variable and therefore, in their phase-space trajectories, even for the canonical Mukhanov-Sasaki field.
	
	In this paper, we investigate whether the degeneracy between Wands-dual backgrounds extends to the realm of real-space quantum information for localized observables.
    We adopt and adapt the Gaussian-state formalism developed in Refs.~\cite{Martin:2021xml, Martin:2021qkg} to construct the covariance matrix of two coarse-grained, spatially localized modes in quasi-de Sitter space. We first examine explicit pairs of Wands-dual backgrounds, such as slow-roll and ultra-slow-roll inflation and two constant-roll realizations, before going on to address the general case of an arbitrary Wands dual pair for a fixed Mukhanov-Sasaki mass. 
	
	Our main result is that, for all such background pairs related by Wands' duality, the symplectic eigenvalues of the two-mode covariance matrix coincide, even though the individual entries of the covariance matrix and the underlying reduced power spectra are markedly different. Since all entropic and correlation measures (entanglement entropy, mutual information, quantum discord, and so on) can be expressed in terms of these symplectic invariants \cite{Serafini:2004,Nicacio:2021, Serafini:book, Cerf:book},	we conclude that Wands-dual backgrounds are indistinguishable for this entire family of localized quantum-informatic probes, at least for Gaussian systems. In fact, even measures such as log negativity, which are determined by the symplectic eigenvalues of the partially-transposed covariance matrix, are shown to be the same for these Wands-dual pairs. In this sense, we uncover a new ``quantum-informatic symmetry'' of quasi-de Sitter vacua: a nontrivial extension of Wands' duality from configuration space power spectra to quantum correlations between local modes of Gaussian fields.
	
	The paper is organized as follows. In Sec-\ref{sec:wands}, we give a precise definition of Wands duals and set up our notation. In Sec-\ref{sec:gaussian}, we describe the real-space coarse-graining of the Gaussian field. In Sec-\ref{sec:symplectic}, we present our main results and show that all entanglement witnesses coincide for Wands-dual pairs. Finally, we conclude in Sec-\ref{sec:conclusions} where we explain the mathematical structure behind this symmetry and outline some potential future directions.

	
	\section{Wands duality and general motivation}
	\label{sec:wands}
	
	A recurring theme in this work is the existence of distinct inflationary histories that share the same linear dynamics for curvature perturbations. This phenomenon is termed as Wands' duality, which states that different background evolutions can yield the same effective mass term in the Mukhanov-Sasaki equation~\cite{Wands:1998yp}. Consequently, the mode equations (for the canonical field) admit the same family of solutions and thus, the corresponding configuration power spectrum coincides across dual backgrounds.
    Our aim is to extend this result to the symplectic invariants of the theory.
	
	To explore this, let us begin by considering a minimally coupled scalar field in a spatially-flat FLRW background $\mathrm{d}s^2 = a^2(\eta) (-\mathrm{d}\eta^2 +  \delta_{ij}\, \mathrm{d}x^i \mathrm{d}x^j)$,
    with the scale factor denoted by $a(\eta)$ where $\eta$ is conformal time. 
    We work with the Mukhanov-Sasaki variable $v = z \zeta$, where $\zeta$ is the comoving curvature perturbation, $z^2 = 2 \epsilon_1 a^2 M_{\mathrm{Pl}}^2$, and $\epsilon_1 = -\dot{H}/H^2$ is the first slow-roll (or Hubble-flow) parameter and $H=\dot{a}/a$ is the Hubble parameter. 
    At leading order in cosmological perturbation theory the action for $v$ is quadratic and takes the form~\cite{Sasaki:1986hm,Mukhanov:1990me}
	\begin{equation}
		S = \frac{1}{2} \int \mathrm{d}^4 x \left[
		\left( v' - \frac{z'}{z} v \right)^2 - (\partial_i v)^2
		\right].
		\label{eq:MS_action}
	\end{equation}
	Varying this action leads to the Mukhanov-Sasaki equation, which in Fourier space is given by
	\begin{equation}
		v_k'' + \left( k^2 - \frac{z''}{z} \right) v_k = 0\,,
		\label{eq:MS_equation}
	\end{equation}
	where the effective mass term \(z''/z\) can be written exactly in terms of the Hubble-flow parameters	\(\epsilon_i\) as \cite{Wands:1998yp}
	\begin{equation}
		\frac{z''}{z}
		= \left( \frac{a'}{a} \right)^2
		\left(
		2 - \epsilon_1 + \frac{3}{2} \epsilon_2 + \frac{1}{4} \epsilon_2^2
		- \frac{1}{2} \epsilon_1 \epsilon_2 + \frac{1}{2} \epsilon_2 \epsilon_3
		\right)\,.
\label{eq:zpp_over_z_general}
	\end{equation}
	Here, $ \epsilon_{i+1} = \mathrm{d}  \ln \epsilon_i / \mathrm{d} N $, where $ N=\ln(a) $ denotes the number of e-folds. Note that ~\eqref{eq:zpp_over_z_general} is exact, and the slow-roll parameters are in general time-dependent, even if, in practice, we often treat them as constant or negligible.
	
	Assuming a quasi-de Sitter background, where $\epsilon_1 \ll 1$,  
    and using the approximate relation $a'/a \simeq - 1/\eta$,\footnote{The conformal time $\eta$ also receives slow--roll corrections, which crucially, are always proportional to $\epsilon_1$. Writing $\eta = \int \mathrm{d}t/a(t) = \int \mathrm{d}N/(aH)$, and using $\mathrm{d}(aH)^{-1}/\mathrm{d}N = -(1+\epsilon_1)/(aH)$, a single integration by parts yields
    \begin{equation*}
        \eta = -\frac{1}{aH (1+\epsilon_1)} - \int \mathrm{d}N \frac{\epsilon_1 \epsilon_2}{aH(1+\epsilon_1)^2}\,,    
    \end{equation*}
    with further iterations producing corrections order by order in slow-roll. In attractor regimes, this expansion is controlled, since $\epsilon_{i>1}$ are genuinely suppressed at higher orders. In non-attractor regimes, $\epsilon_2$ must be replaced by a more appropriate, genuine small parameter (the choice is model dependent; see e.g.~\cite{Pattison:2019hef}). However, an $\epsilon_1$ factor is inherited at every order, which justifies the leading-order approximation $a'/a \simeq -1/\eta$ even in non-attractor backgrounds.}
    one can rewrite the effective mass in the form 
	\begin{equation}
		\frac{z''}{z} \simeq \frac{1}{\eta^2}
		\left[
		2 + \frac{3}{2} \epsilon_2 + \frac{1}{4} \epsilon_2^2
		+ \mathcal{O}(\epsilon_1, \epsilon_3)
		\right]\,.
		\label{eq:zpp_over_z_slowroll}
	\end{equation}
	This expression plays a central role in understanding the mathematical structure behind the duality. It is convenient to parametrize this as
	\begin{equation}
		\frac{z''}{z} = \frac{\nu^2 - 1/4}{\eta^2}\,,
		\label{eq:zpp_over_z_nu}
	\end{equation}
	such that \eqref{eq:MS_equation} becomes 
	\begin{equation}
		v_k''(\eta) + \left( k^2 - \frac{\nu^2 - 1/4}{\eta^2} \right) v_k(\eta) = 0\,.
		\label{eq:MS_Bessel}
	\end{equation}
	Assuming $\nu$ to be constant, the solutions corresponding to the Bunch-Davies vacuum are given by
	\begin{equation}
		v_k(\eta) = 
		\frac{1}{2} e^{i \frac{\pi}{2} \left(\nu + \frac{1}{2}\right)}
		\sqrt{-\pi \eta} \, H^{(1)}_{\nu}(-k\eta)\,,
		\label{eq:mode_Hankel}
	\end{equation}
	where $H^{(1)}_{\nu}$ denotes a Hankel function of the first kind and
	\begin{equation}
		\nu^2 = \frac{9}{4} + \frac{3}{2} \epsilon_2
		+ \frac{1}{4} \epsilon_2^2 + \mathcal{O}(\epsilon_1, \epsilon_3)\,.
		\label{eq:nu_slowroll}
	\end{equation}
	The key observation here is that different combinations of the Hubble-flow parameters $\epsilon_i$ can produce the same value of $\nu$. For explicitness, in what follows we restrict to backgrounds in which $\epsilon_2$ is the only non-negligible Hubble-flow parameter, in which case~\eqref{eq:nu_slowroll} reduces to the perfect square $\nu^2 = (\epsilon_2 + 3)^2/4$ and the Wands duality transformation acts as the simple mapping $\epsilon_2 \to -6 - \epsilon_2$.\footnote{In more general non-attractor settings, $\nu$ can remain constant in time while $\epsilon_2$ varies, with compensating contributions involving $\epsilon_3$ preventing $\nu$ from changing. This is the case, for instance, in the USR-to-SR relaxation discussed in~\cite{Briaud:2025hra}.} The argument we develop in the following sections is, however, insensitive to this restriction, and applies to any pair of backgrounds related by a transformation of the Hubble-flow parameters that leaves $\nu$ invariant.
    Consequently, the corresponding distinct pair of backgrounds lead to identical mode equations which, in turn, leads to exactly identical families of solutions for the Mukhanov-Sasaki field $v_k$, given Bunch-Davies initial conditions.
	
	A canonical example of this is the duality between the pair formed by slow-roll (SR) and ultra slow-roll (USR) inflation. In the SR case, all $\epsilon_i$ parameters in \eqref{eq:nu_slowroll} are small and can be neglected to leading order, which effectively	describes the limit that the curvature perturbations are massless. In contrast, USR inflation is characterized by $\epsilon_2 = -6$ while the remaining Hubble flow parameters can be considered as negligible. As a result, the squeezing term becomes $ z''/z \approx 2/\eta^2 $ in both scenarios, leading to identical mode equations and, hence, identical solutions characterized by $\nu \approx 3/2$. For both cases, \eqref{eq:mode_Hankel} becomes
	\begin{equation}
		v^{\mathrm{SR}}_k(\eta) = v^{\mathrm{USR}}_k(\eta)
		= \frac{\mathrm{e}^{-i k \eta}}{\sqrt{2k}}
		\left( 1 - \frac{i}{k\eta} \right)\,.
		\label{eq:SRUSR_modes}
	\end{equation}
	
	Although the mode functions coincide,  discrepancies arise between the corresponding conjugate momenta, namely that 
	\begin{equation}
		\pi_k(\eta) = v_k' - \frac{z'}{z} v_k = v_k' - \frac{a'}{a}
		\left( 1 + \frac{\epsilon_2}{2} \right) v_k\,,
		\label{eq:pi_general}
	\end{equation}
	remains sensitive to the background even under the same initial conditions. Thus, this equivalence does not imply that the phase-space dynamics of the perturbations are identical between the two duals.
	For SR, one finds
	\begin{equation}
		\pi^{\mathrm{SR}}_k(\eta) = - i \sqrt{\frac{k}{2}} \, \mathrm{e}^{-i k \eta}\,,
	\end{equation}
	while for USR, we have 
	\begin{equation}
		\pi^{\mathrm{USR}}_k(\eta) = - i \sqrt{\frac{k}{2}} \, \mathrm{e}^{-i k \eta}
		\left[ 1 - \frac{3i}{k\eta} - \frac{3}{(k\eta)^2} \right]\,.
	\end{equation}
	Thus, even at the classical level, the phase-space trajectories corresponding to Wands-dual backgrounds differ, despite sharing the same $v_k$ and power spectrum for the Mukanov-Sasaki field\footnote{Another way to think about this is that the physical curvature perturbation $\zeta$ is different for SR and USR even if the canonical Mukhanov-Sasaki field is the same. As we shall show in the end, this is the crucial physical difference between these Wands-dual pairs.}. In what follows, we shall demonstrate that this difference in phase space does not propagate to the symplectic invariants of the localized bipartite system we construct, which is the origin of the new quantum-informatic symmetry that we find.


	\section{Coarse-grained perturbations of a Gaussian field}
	\label{sec:gaussian}
	
	
	\subsection{Quantization and two-point functions}
	
	To examine the entanglement structure of a theory, it is essential for us to transition from Fourier to real space. This is so that we can localize the quantum field to finite volumes in space and find the entanglement between these regions. To quantify quantum correlations in real space, we adapt the continuous-variable formalism introduced in Refs.~\cite{Martin:2021qkg,Martin:2021xml} to the Mukhanov-Sasaki field. In this framework, two-point correlation functions evaluated at spatial locations ${\bf x_1}$ and ${\bf x_2}$ can be described by a Gaussian bipartite system which, in turn, is fully characterized by its covariance matrix.
	
	We collect the field and its conjugate momentum into a two-component phase-space vector $\hat{q} = (\hat{v}, \hat{\pi})$, and expand each component in Fourier modes as
	\begin{equation}
		\hat{q}_i(\mathbf{x}) =
		\int \frac{\mathrm{d}^3 k}{(2\pi)^{3/2}} \;
		\hat{q}_i(\mathbf{k}) \, \mathrm{e}^{i \mathbf{k}\cdot\mathbf{x}}\,,
	\end{equation}
	with the reality condition \(\hat{q}_i(\mathbf{k}) = \hat{q}_i^\dagger(-\mathbf{k})\)\,.
	
	The canonical commutation relations, in both real and Fourier space, can be written compactly as
	\begin{equation}\label{eq:CCR}
		\big[ \hat{q}_i(\mathbf{x}_1), \hat{q}_j(\mathbf{x}_2) \big] 
		= i\, J^{(1)}_{ij} \, \delta(\mathbf{x}_1 - \mathbf{x}_2)\,,
		\qquad
		\big[ \hat{q}_i(\mathbf{k}_1), \hat{q}_j^\dagger(\mathbf{k}_2) \big] 
		= i\, J^{(1)}_{ij} \, \delta(\mathbf{k}_1 - \mathbf{k}_2)\,,
	\end{equation}
where the $2 \times 2$ symplectic matrix, given by
	\begin{equation}
		{\bf J}^{(1)} =
		\begin{pmatrix}
			0 & 1 \\
			-1 & 0
		\end{pmatrix}\,,
	\end{equation}
	encodes the usual position-momentum structure.
	
	For Gaussian states, all the statistical information is contained in the two-point correlators, which is most conveniently expressed in terms of the expectation values of the anticommutators of the fields (since the commutator gives a state-independent result), namely
	\begin{align}
		\big\langle \{ \hat{q}_i(\mathbf{x}_1), \hat{q}_j(\mathbf{x}_2) \} \big\rangle
		&= \int \frac{\mathrm{d}^3k_1}{(2\pi)^{3/2}} \frac{\mathrm{d}^3k_2}{(2\pi)^{3/2}}
		\, \mathrm{e}^{i (\mathbf{k}_1\cdot\mathbf{x}_1 - \mathbf{k}_2\cdot\mathbf{x}_2)}
		\big\langle \{ \hat{q}_i^\dagger(\mathbf{k}_1), \hat{q}_j(\mathbf{k}_2) \} \big\rangle\,,
	\end{align}
	where $\{A,B\} = (AB + BA)/2$. Assuming translation invariance,     the Fourier-space anticommutators can be parametrized as
	\begin{equation}
		\big\langle \{ \hat{q}_i^\dagger(\mathbf{k}_1), \hat{q}_j(\mathbf{k}_2) \} \big\rangle
		= \frac{2\pi^2}{k_1^3} {\cal P}_{ij} ({\bf k}_1) \, \delta(\mathbf{k}_1-\mathbf{k}_2)\,,
	\end{equation}
which defines the reduced power spectra ${\cal P}_{ij} ({\bf k}_1)$. Further, invariance under rotations naturally implies ${\cal P}_{ij} ({\bf k}) = {\cal P}_{ij} (k)$, which leads to the real-space anticommutators having the form
	\begin{equation}
		\big\langle \{ \hat{q}_i(\mathbf{x}_1), \hat{q}_j(\mathbf{x}_2) \} \big\rangle
		= \int_0^{\infty} \mathrm{d}\ln k \, {\cal P}_{ij}(k) \,
		\mathrm{sinc}\!\left( k |\mathbf{x}_1 - \mathbf{x}_2| \right) ,
	\end{equation}
	where $\mathrm{sinc}\,\Theta \equiv \sin\Theta / \Theta$.
	
	
	\subsection{Coarse-graining and canonical structure}
	
	Realistic measurements are sensitive only to field values smoothed over finite physical scales. 
    We model this via the convolution of the field with a window function $W$ with characteristic physical scale $R$, and define the coarse-grained field operators as:
	\begin{equation}
		\hat{q}_{R,i}(\mathbf{x})
		\equiv \left( \frac{a}{R} \right)^3
		\int \mathrm{d}^3\mathbf{y} \,
		W\!\left( \frac{a|\mathbf{y} - \mathbf{x}|}{R} \right)
		\hat{q}_i(\mathbf{y})\,,
	\end{equation}
	where \(\mathbf{x}\) and \(\mathbf{y}\) are comoving coordinates. Note that all ratios involve physical distances via the scale factor $a$, ensuring consistency within the expanding background. The normalization of $W$ is chosen such that a uniform field configuration is left invariant after the coarse-graining, \textit{i.e.,}
	\begin{equation}
        \label{eq:normalisation}
		4\pi \int_0^{\infty} \mathrm{d}x \, x^2 W(x) = 1\,.
	\end{equation}
	Moreover, we require 
    that $W(x) \simeq 0$ for $x \gg 1$, so that only field values at points within a physical distance $\sim \mathcal{O}(R)$ of $\mathbf{x}$ contribute significantly to $\hat{q}_{R,i}(\mathbf{x})$.
		
	In Fourier space, coarse-graining corresponds to multiplying each mode by the (spherically symmetric)	transform\footnote{Explicitly, this is done through a simple change of variables ${\bf u} = k({\bf x}-{\bf y})$, and integrating over ${\bf y}$ and the angular components of ${\bf u}$.} \(\widetilde{W}\),
	\begin{equation}
		\hat{q}_{R,i}(\mathbf{k}) = \hat{q}_i(\mathbf{k}) \,
		\widetilde{W}\!\left( \frac{kR}{a} \right)\,,
	\end{equation}
	where
	\begin{equation}
		\widetilde{W}\!\left( \frac{kR}{a} \right)
		= 4\pi \left( \frac{a}{kR} \right)^3 \!
		\int_0^{\infty} \mathrm{d}u \, u \sin u \,
		W\!\left( \frac{a u}{kR} \right)\,.
	\end{equation}
    A sharp top-hat, such as $W(x) = 3/(4\pi)\  \theta(1-x)$, satisfies the normalisation condition~\eqref{eq:normalisation}, which in turn implies $\widetilde{W}(0) = 1$. Moreover, $\widetilde{W}(kR/a) \simeq 1$ for $kR/a \ll 1$, signalling that long-wavelength modes pass through the filter nearly unaffected, as would be expected. However, the sharp cutoff at $x = 1$ causes $\widetilde{W}$ to decay only as $(kR/a)^{-2}$ for $kR/a \gg 1$, which leads to logarithmically UV-divergent integrals for a scale-invariant spectrum.
    Following Ref.~\cite{Martin:2021qkg}, we therefore adopt the mildly-smoothed window function of ~\eqref{eq:window-function}, controlled by a parameter $\delta$, for which $\widetilde{W}$ and the equal-point commutator are both analytically tractable; these details have been summarized in Appendix~\ref{app:window}.
	
	With this choice, the commutator of two coarse-grained operators at equal spatial point, reads
	\begin{equation}
		\big[ \hat{q}_{R,1}(\mathbf{x}), \hat{q}_{R,2}(\mathbf{x}) \big]
		= i \, \frac{3}{4\pi} \left( \frac{a}{R} \right)^3 G(\delta)\,,
	\end{equation}
	where the dimensionless function \(G(\delta)\) is obtained by integrating \(W^2\) over space. 
    This clearly shows that the coarse-graining prescription does not automatically yield the canonical commutation relations even though the commutators between fields of the same type are still trivially satisfied. However, more generally in this case, for our smoother window function, this is satisfied for $a |{\bf x}_1 - {\bf x}_2| > 2R (1+\delta)$. We will proceed under this constraint, and then recover the standard structure of \eqref{eq:CCR} in the coincident limit ${\bf x}_1 = {\bf x}_2$.
	
	To recover the canonical structure and eliminate the prefactor, we perform a local rescaling of the field variables as
	\begin{equation}
    \label{eq:rescaling}
		\tilde{q}_R = \Lambda^{(1)} \hat{q}_R\,,
		\qquad 
		\Lambda^{(1)} \equiv
		\left( \frac{R}{a} \right)^{3/2}
		\left( \frac{4\pi}{3 G(\delta)} \right)^{1/2}
		\begin{pmatrix}
			\lambda & 0 \\
			0 & \lambda^{-1}
		\end{pmatrix}\,,
	\end{equation}
	where $\lambda$ is a dimensionless parameter ensuring that the two phase-space components have	the same physical units and which will be convenient for tracking local symplectic transformations. It also serves as a useful bookkeeping device for verifying the invariance of local symplectic structures, since this transformation corresponds to a phase-space dilatation. With this choice, the rescaled coarse-grained variables satisfy
	\begin{equation}
		\big[ \tilde{q}_{R,1}(\mathbf{x}), \tilde{q}_{R,2}(\mathbf{x}) \big] = i\,.
	\end{equation}
	
	
	\subsection{Bipartite system and covariance matrix}
	
	To extract any meaningful information in this framework, we must compare field data at two or more spatial points. To study spatial correlations we restrict our attention to just two coarse-grained regions, centred at comoving positions $\mathbf{x}_1$ and $\mathbf{x}_2$. We begin by collecting the corresponding rescaled variables into a four-component phase-space vector
	\begin{equation}
		\tilde{Q}_R(\mathbf{x}_1,\mathbf{x}_2) =
		\begin{pmatrix}
			\tilde{v}_R(\mathbf{x}_1) \\
			\tilde{\pi}_R(\mathbf{x}_1) \\
			\tilde{v}_R(\mathbf{x}_2) \\
			\tilde{\pi}_R(\mathbf{x}_2)
		\end{pmatrix}\,,
	\end{equation}
	obtained from the unrescaled vector by a block-diagonal transformation\footnote{$\tilde{Q}_R$ is the column vector collecting all of the $\tilde{q}_R$ elements from the previous subsection.}
	\begin{equation}
		\tilde{Q}_R(\mathbf{x}_1,\mathbf{x}_2)
		= {\bf\Lambda}^{(2)} \hat{Q}_R(\mathbf{x}_1,\mathbf{x}_2)\,,
	\end{equation}
	where the transformation matrix is given by
	\begin{equation}
		{\bf \Lambda}^{(2)} = {\bf \Lambda}^{(1)} \oplus {\bf \Lambda}^{(1)} = \left(\frac{R}{a}\right)^{3/2} 
		\sqrt{\frac{4\pi}{3G(\delta)}} 
		\begin{pmatrix}
			\lambda & 0 & 0 & 0 \\
			0  & \lambda^{-1} & 0 & 0 \\
			0 & 0 & \lambda & 0 \\
			0 & 0 & 0 & \lambda^{-1}
		\end{pmatrix}\,.
	\end{equation}
	
	As a result,the commutation relations now take the canonical form
	\begin{equation}
    \label{eq:canonical:relations:rescaled}
		\big[ \tilde{Q}_{R,a}(\mathbf{x}_1,\mathbf{x}_2),
		\tilde{Q}_{R,b}(\mathbf{x}_1,\mathbf{x}_2) \big]
		= i J^{(2)}_{ab}\,,
		\quad {\rm where} \quad {\bf J}^{(2)} = {\bf J}^{(1)} \oplus {\bf J}^{(1)}\,.
	\end{equation}
	
	Equipped with these tools, we can now define the covariance matrix $\boldsymbol{\gamma}$, which encodes all the statistical information of the resulting bipartite system, from the two-point functions of the rescaled phase-space variables as 
	\begin{equation}
		\left\langle \{ \tilde{Q}_{R,a}, \tilde{Q}_{R,b} \} \right\rangle
		=  \gamma_{ab}\,.
	\end{equation}
	Using the expressions above for the coarse-grained fields in Fourier space, one finds the explicit form
	\begin{equation}
		{\bf \gamma} = 
		\frac{8\pi}{3 G(\delta)}
		\left( \frac{R}{a} \right)^3
		\int \mathrm{d}\ln k \,
		\widetilde{W}^2\!\left( \frac{kR}{a} \right)
		\Gamma(k,d)\,,
		\label{eq:gamma_general}
	\end{equation}
	where $d \equiv a |\mathbf{x}_1 - \mathbf{x}_2|$ is the physical separation between the two patches, and
	\begin{equation}
		\Gamma(k,d) =
		\begin{pmatrix}
			\lambda^2 P_{vv}(k) & P_{v\pi}(k) & \lambda^2 P_{vv}(k) \, \mathrm{sinc}\!\left( \frac{k d}{a} \right) & P_{v\pi}(k) \, \mathrm{sinc}\!\left( \frac{k d}{a} \right) \\
			P_{v\pi}(k) & \lambda^{-2} P_{\pi\pi}(k) & P_{v\pi}(k) \, \mathrm{sinc}\!\left( \frac{k d}{a} \right) & \lambda^{-2} P_{\pi\pi}(k) \, \mathrm{sinc}\!\left( \frac{k d}{a} \right) \\
			\lambda^2 P_{vv}(k) \, \mathrm{sinc}\!\left( \frac{k d}{a} \right) & P_{v\pi}(k) \, \mathrm{sinc}\!\left( \frac{k d}{a} \right) & \lambda^2 P_{vv}(k) & P_{v\pi}(k) \\
			P_{v\pi}(k) \, \mathrm{sinc}\!\left( \frac{k d}{a} \right) & \lambda^{-2} P_{\pi\pi}(k) \, \mathrm{sinc}\!\left( \frac{k d}{a} \right) & P_{v\pi}(k) & \lambda^{-2} P_{\pi\pi}(k)
		\end{pmatrix}\,.
	\end{equation}
	By construction, ${\bf \gamma}$ is symmetric, and is also invariant under the exchange \(\mathbf{x}_1 \leftrightarrow \mathbf{x}_2\),	so that it is fully characterized by six independent entries, which we may choose to be $\gamma_{11}, \gamma_{12}, \gamma_{13}, \gamma_{14}, \gamma_{22}, \gamma_{24}$.

	
	\section{Symplectic invariants and Wands duals}
	\label{sec:symplectic}

    The statistical properties characterizing a Gaussian quantum field are fully encoded in its covariance matrix, which in turn is built from the two-point correlators of the quadrature operators (e.g., $v$ and $\pi$). Then, it is natural to ask what properties of the covariance matrix are physical in the sense of being independent of the choice of field basis or canonical frame. Symplectic invariants/eigenvalues are central in answering this question. 

    In a phase-space description, the natural transformations that preserve the canonical structure are symplectic transformations $S \in \mathrm{Sp}(2N, \mathbb{R})$, which satisfy $S J S^T = J$, and under which the covariance matrix transforms as $\gamma \to S \gamma S^T$ \cite{Weedbrook:2011wxo}. A classical result by Williamson~\cite{Williamson1936} states that one can always find a simplectic transformation that brings any (positive) covariance matrix into a diagonal form. The entries of such a matrix, denoted as $\{\sigma_k\}$, are precisely the symplectic eigenvalues. In practice, they are computed as the {\it standard} eigenvalues of the matrix $|i J \gamma|$. This eigenspectrum is unique and thus invariant under all symplectic transformations, encoding the fundamental properties of the Gaussian quantum state. Case in point, the uncertainty principle bounds the elements $\sigma_k$ from below as $\sigma_k\geq 1$, with pure states saturating the bound. In this section, we shall construct the invariants $\{\sigma_k\}$, and other quantities derived from them, in a  way that is naturally suited to the framework introduced in previous sections.
	
	\subsection{Symplectic spectrum and auxiliary integrals}
	
	In the bipartite setting described above, the covariance matrix $\gamma$ has two nontrivial symplectic eigenvalues, denoted $\sigma_{\pm}$. They are defined as the positive eigenvalues of the matrix $i {\bf J}^{(2)} {\bf \gamma}$, and can be expressed directly in terms of the position-momentum blocks of ${\bf \gamma}$ as \cite{Serafini:book,Weedbrook:2011wxo,Martin:2021qkg}
	\begin{equation}
		\sigma_{\pm}
		= \sqrt{ (\gamma_{11} \pm \gamma_{13})(\gamma_{22} \pm \gamma_{24})
			- (\gamma_{12} \pm \gamma_{14})^2 }\,.
		\label{eq:sigma_pm}
	\end{equation}
	It is also convenient to introduce the single-point invariant
	\begin{equation}
		\sigma_1 = \sigma_2 = \sqrt{\det \gamma_1}
		= \sqrt{\gamma_{11} \gamma_{22} - \gamma_{12}^2}\,,
		\label{eq:sigma1}
	\end{equation}
	where ${\bf \gamma}_1$ is the covariance matrix obtained by `tracing over' the degrees of freedom localized at \(\mathbf{x}_2\), as well as the inter-patch invariant
	\begin{equation}
		\sigma_{1-2} = \sqrt{\det \gamma_{1-2}}
		\equiv \sqrt{\gamma_{13} \gamma_{24} - \gamma_{14}^2}\,.
		\label{eq:sigma12}
	\end{equation}
	These quantities are invariant under local symplectic transformations acting separately on the two patches and therefore serve as building blocks for Gaussian entanglement and correlation measures \cite{Serafini:2004,Weedbrook:2011wxo,Wang:2007}. 
	
	To evaluate the entries of ${\bf \gamma}$ in specific backgrounds, it is useful to factor out the dependence on the power spectra into a small set of auxiliary integrals. Let us introduce the dimensionless wavenumber $w \equiv k R/a$ and an effective IR cutoff $\beta$ defined by
	\begin{equation}
		\beta \equiv \frac{R}{R_{\text{obs}}} < 1\,,
		\qquad
		R_{\text{obs}} = \frac{\mathrm{e}^{N_{\text{inf}}}}{H}\,,
	\end{equation}
	where $R_{\rm obs}$ represents the size of the observable Universe, $H$ is the Hubble parameter during inflation, and $N_{\rm inf}$ is the number of $e$-folds the largest observable modes spent outside the horizon. Using these, and recalling $\delta$ as the parameter associated with smoothening the window function, we define the auxiliary integrals as
	\begin{equation}
		K(\beta, \mu, \delta)
		\equiv \int_{\beta}^{\infty} \mathrm{d}w \, w^{\mu} \;
		\widetilde{W}^2(w)\,,
		\qquad
		L(\beta, \mu, \delta, \alpha)
		\equiv \int_{\beta}^{\infty} \mathrm{d}w \ w^{\mu} \;
		\widetilde{W}^2(w) \, \mathrm{sinc}(\alpha w)\,,
		\label{eq:K_L_def}
	\end{equation}
	where $\alpha \equiv d/R$ is the dimensionless separation parameter between the patches, which is constrained to be $\alpha > 2(1+\delta)$ and $\alpha \beta \ll 1$, so that the two regions are widely separated compared to the smoothing scale and are yet both well within the observable Universe \cite{Martin:2021qkg}.
	
	
\subsection{The slow-roll/ultra-slow-roll pair}
\label{subsec:SR_USR}

	We first recap the SR case, which was first studied in Ref.~\cite{Martin:2021qkg}, below. Using the mode functions in ~\eqref{eq:SRUSR_modes} and the corresponding conjugate momenta,	the reduced power spectra for a massless field in de Sitter read 
	\begin{equation}
		P_{vv}(k) = \frac{1 + k^2 \eta^2}{4\pi^2 \eta^2}\, ,
		\qquad
		P_{v\pi}(k) = \frac{k^2}{4\pi^2 \eta}\, ,
		\qquad
		P_{\pi\pi}(k) = \frac{k^4}{4\pi^2}\, .
		\label{eq:SR_power}
	\end{equation}
	Substituting these expressions into ~\eqref{eq:gamma_general}, we obtain the following form for the covariance matrix elements in the SR regime~\cite{Martin:2021qkg}
	\begin{align}
		\gamma^{\mathrm{SR}}_{11} &=
		\frac{2\lambda^2}{3\pi G(\delta)}(HR)^2 \left(\frac{R}{a}\right)
		\left[ K(\beta,-1,\delta) + \frac{1}{(HR)^2}K(\beta,1,\delta) \right]\,, \\
		\gamma^{\mathrm{SR}}_{12} &= - \frac{2}{3\pi G(\delta)}(HR)K(\beta,1,\delta)\,, \\
		\gamma^{\mathrm{SR}}_{22} &=
		\frac{2}{3\pi\lambda^2 G(\delta)}\left(\frac{R}{a}\right)^{-1}K(\beta,3,\delta)\,, \\
		\gamma^{\mathrm{SR}}_{13} &=
		\frac{2\lambda^2}{3\pi G(\delta)}(HR)^2 \left(\frac{R}{a}\right)
		\left[ L(\beta,-1,\delta,\alpha) + \frac{1}{(HR)^2}L(\beta,1,\delta,\alpha) \right]\,, \\
		\gamma^{\mathrm{SR}}_{14} &= - \frac{2}{3\pi G(\delta)}(HR)L(\beta,1,\delta,\alpha)\,, \\
		\gamma^{\mathrm{SR}}_{24} &=
		\frac{2}{3\pi\lambda^2 G(\delta)}\left(\frac{R}{a}\right)^{-1}L(\beta,3,\delta,\alpha)\,.
		\label{eq:gamma_SR_all}
	\end{align}
	
\begin{figure}
	    \centering
	    \includegraphics[width=0.9\linewidth]{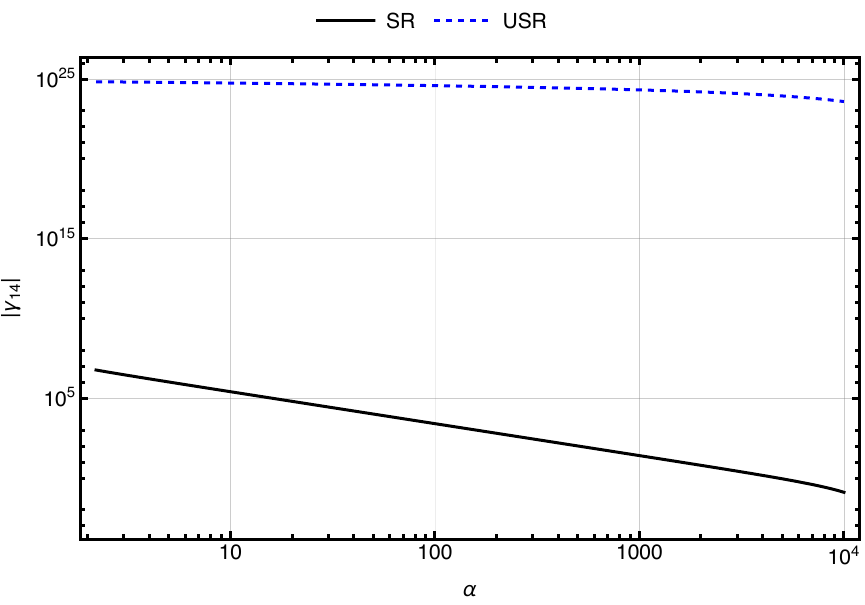}
	    \caption{Component $|\gamma_{14}|$ of the covariance matrix in SR and USR for $\beta=10^{-4}$, $H R=10^8$ and $\delta=0.1$. The value of $\gamma_{14}$ is negative for SR and positive for USR. This illustrates that the covariance matrix is different in these two backgrounds. The inset panel zooms in on the SR behaviour.
        }
	    \label{fig:sub1}
	\end{figure}

We now repeat the analysis for a phase of USR inflation\footnote{Phenomenologically speaking, this is not how one would expect a period of USR inflation to take place. In other words, USR can never be a standalone background since it can only be a transient regime. However, this point is not important for the general purpose of this work.} with Bunch-Davies initial conditions. As before, we begin with the relevant components of the reduced power spectrum:
	\begin{equation}
		P_{vv}(k) = \frac{1 + k^2 \eta^2}{4\pi^2 \eta^2}\,,
		\qquad
		P_{v\pi}(k) = \frac{-3 + 2k^2 \eta^2}{4\pi^2 \eta^3}\,,
		\qquad
		P_{\pi\pi}(k) = \frac{9 + 3k^2 \eta^2 + k^4 \eta^4}{4\pi^2 \eta^4}\,.
		\label{eq:USR_power}
	\end{equation}
	These expressions are then inserted into the general formula for the covariance matrix~\eqref{eq:gamma_general}, yielding the following components for the bipartite system in the USR	background as 
	\begin{align}
		\gamma^{\mathrm{USR}}_{11} &=
		\frac{2\lambda^2}{3\pi G(\delta)}(HR)^2 \left(\frac{R}{a}\right)
		\left[ K(\beta,-1,\delta) + \frac{1}{(HR)^2}K(\beta,1,\delta) \right]\,, \\
		\gamma^{\mathrm{USR}}_{12} &=
		\frac{2}{3\pi G(\delta)}\left[ 2HR\,K(\beta,1,\delta) + 3H^3R^3K(\beta,-1,\delta) \right]\,, \\
		\gamma^{\mathrm{USR}}_{22} &=
		\frac{2}{3\pi\lambda^2 G(\delta)}\left(\frac{R}{a}\right)^{-1}
		\left[ K(\beta,3,\delta) + 9H^4R^4K(\beta,-1,\delta) + 3H^2R^2K(\beta,1,\delta) \right]\,, \\
		\gamma^{\mathrm{USR}}_{13} &=
		\frac{2\lambda^2}{3\pi G(\delta)}(HR)^2 \left(\frac{R}{a}\right)
		\left[ L(\beta,-1,\delta,\alpha) + \frac{1}{(HR)^2}L(\beta,1,\delta,\alpha) \right]\,, \\
		\gamma^{\mathrm{USR}}_{14} &=
		\frac{2}{3\pi G(\delta)}\left[ 2HR\,L(\beta,1,\delta,\alpha)
		+ 3H^3R^3L(\beta,-1,\delta,\alpha) \right]\,, \\
		\gamma^{\mathrm{USR}}_{24} &=
		\frac{2}{3\pi\lambda^2 G(\delta)}\left(\frac{R}{a}\right)^{-1}
		\left[ L(\beta,3,\delta,\alpha) + 9H^4R^4L(\beta,-1,\delta,\alpha)
		+ 3H^2R^2L(\beta,1,\delta,\alpha) \right]\,.
		\label{eq:gamma_USR_all}
	\end{align}
See Fig.~\ref{fig:sub1} as an illustration that the components of the covariance matrix for SR and USR are different.

	To make the equality of the symplectic eigenvalues explicit, it is convenient to introduce the combinations
	\begin{equation}
		\mathcal{K}_{m}^{(\pm)} \equiv
		K(\beta,m,\delta) \pm L(\beta,m,\delta,\alpha),
		\qquad m \in \{-1,1,3\}.
	\end{equation}
	Using Eqs.~\eqref{eq:gamma_SR_all} in the definition~\eqref{eq:sigma_pm}, one finds after factoring out the common prefactors that the squared symplectic eigenvalues in the SR case can be written as
	\begin{equation}
		\left(\sigma_\pm^{\mathrm{SR}}\right)^2
		= \frac{4(HR)^2}{9\pi^2 G(\delta)^2}
		\left\{
		\left[ \mathcal{K}_{-1}^{(\pm)} + \frac{1}{(HR)^2}\mathcal{K}_{1}^{(\pm)} \right]
		\mathcal{K}_{3}^{(\pm)}
		- \left( \mathcal{K}_{1}^{(\pm)} \right)^2
		\right\}.
		\label{eq:sigma_SR_explicit}
	\end{equation}
	Repeating the calculation with the USR entries in ~\eqref{eq:gamma_USR_all}, one finds that all additional terms proportional to $H^2R^2$ and $H^4R^4$ cancel out in the combination $(\gamma_{11}\pm\gamma_{13})(\gamma_{22}\pm\gamma_{24})-(\gamma_{12}\pm\gamma_{14})^2$, so that the squared symplectic eigenvalues in the USR case can be written in exactly the same form:
	\begin{equation}
		\left(\sigma_\pm^{\mathrm{USR}}\right)^2
		= \frac{4(HR)^2}{9\pi^2 G(\delta)^2}
		\left\{
		\left[ \mathcal{K}_{-1}^{(\pm)} + \frac{1}{(HR)^2}\mathcal{K}_{1}^{(\pm)} \right]
		\mathcal{K}_{3}^{(\pm)}
		- \left( \mathcal{K}_{1}^{(\pm)} \right)^2
		\right\}.
		\label{eq:sigma_USR_explicit}
	\end{equation}
	In particular,
	\begin{equation}
		\left(\sigma_\pm^{\mathrm{SR}}\right)^2
		= \left(\sigma_\pm^{\mathrm{USR}}\right)^2\, ,
	\end{equation}
	so that the symplectic eigenvalues, and hence all quantum entanglement and correlation measures built from them, are identical for the SR/USR Wands-dual pair.

	\subsection{General Wands duals for constant-roll}
	\label{subsec:general_duals}
    The SR/USR pair discussed above motivates a broader class of Wands-dual backgrounds. In what follows, we consider
    more general classes which are defined by a time-independent Hankel index $\nu$, and which we refer to as 
    constant-roll (CR) backgrounds. Under the approximation adopted in Section~\ref{sec:wands}, where only the 
    Hubble-flow parameter $\epsilon_2$ is retained, constant $\nu$ is equivalent to constant $\epsilon_2$.
    Then, inverting the relation 
    $\nu^2 = (\epsilon_2 + 3)^2/4$ yields the two values of $\epsilon_2$ associated with a given $\nu$,
    \begin{equation}\label{wands}
    	\epsilon_2 = -3 \pm 2\nu\,,
    \end{equation}
    each corresponding to one member of the Wands-dual pair. The standard case $\nu = 3/2$ recovers SR/USR with 
    $\epsilon_2 = 0$ and $-6$, respectively. As a representative CR example beyond SR/USR, $\nu = 5/2$ yields the dual 
    pair $\epsilon_2 = 2$ and $-8$ (see Appendix~\ref{app:explicit_sigmas}).

    The configuration-field mode functions are still given by~\eqref{eq:mode_Hankel}, and are common to both members of 
    the pair. On the contrary, the conjugate momenta~\eqref{eq:pi_general} differ between the duals. Using the recurrence 
    relations of the Hankel functions, they can be written compactly for general (constant) $\nu$ as
    \begin{align}
    	\pi_k^{(I)}(\eta) &= -\frac{1}{2}\sqrt{k\pi}\,
    	e^{i\frac{\pi}{2}(\nu+1/2)}(-k\eta)H^{(1)}_{\nu-1}(-k\eta)\,, \\
    	\pi_k^{(II)}(\eta) &= \frac{1}{2}\sqrt{k\pi}\,
    	e^{i\frac{\pi}{2}(\nu+1/2)}(-k\eta)H^{(1)}_{\nu+1}(-k\eta)\,,
    \end{align}
    where the $\left(I, II\right)$ labels correspond to the two solutions of~\eqref{wands}.
	From these, one obtains two distinct sets of reduced power spectra $P^{\mathrm{(I)}}_{ij}$ and $P^{\mathrm{(II)}}_{ij}$, and hence two different covariance matrices $\gamma^{\mathrm{(I)}}$ and $\gamma^{\mathrm{(II)}}$.	Nevertheless, the symplectic eigenvalues computed from ~\eqref{eq:sigma_pm} are again identical for the	dual pair, and so are all quantum correlation measures constructed from them. In Appendix~\ref{app:general_duals}, we show this explicitly using the above expressions.

\begin{figure}[h]
    \centering

    \begin{subfigure}{0.48\linewidth}
        \centering
        \includegraphics[width=\linewidth]{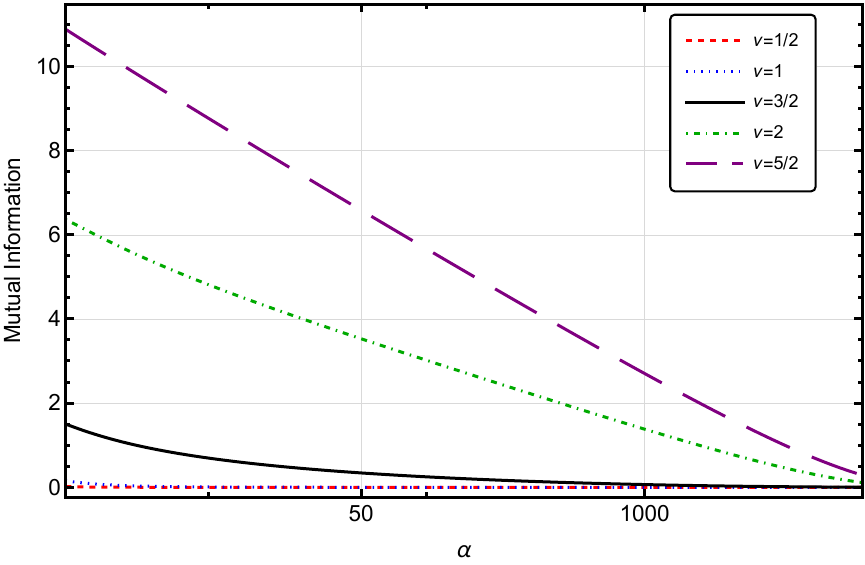}
        \caption{Mutual information in linear scale}
    \end{subfigure}
    \hfill
    \begin{subfigure}{0.48\linewidth}
        \centering
        \includegraphics[width=\linewidth]{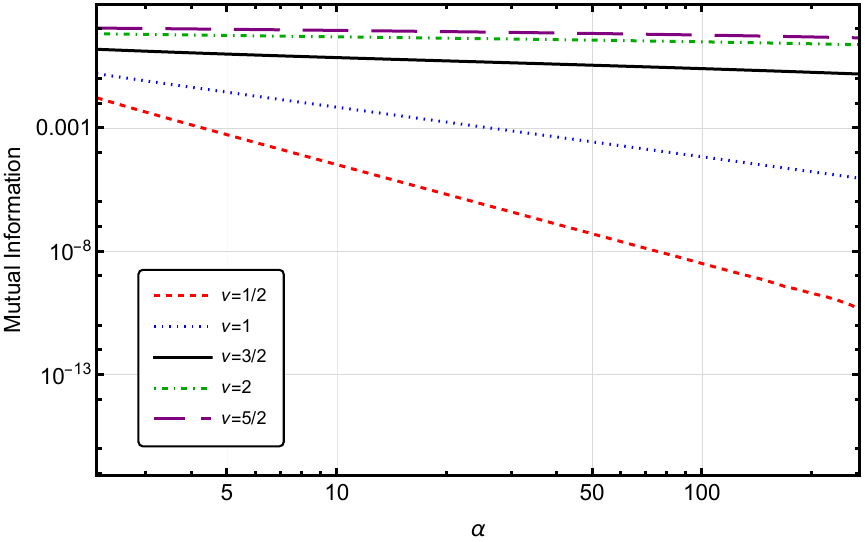}
        \caption{Mutual information in log scale}
    \end{subfigure}

    \caption{
    Mutual information as a function of $\alpha$ for different Wands duals. The left panel uses linear scale while the right 
    panel uses logarithmic scale.  As before, we have set $\beta=10^{-4}, HR=10^8$ and $\delta=0.1$.  The mutual information decreases with the distance $\alpha$ between the two patches for all $\nu$ considered. On the other hand, as $\nu$ increases, the mutual information also increases.
    }

    \label{fig:placeholder1}
\end{figure}

    Figure \ref{fig:placeholder1} illustrates the mutual information behaviour for different Wands dual pairs. The mutual information of a Gaussian bipartite is defined as:
    \begin{equation}
\mathcal{I}=S\left(\rho_1\right)+S\left(\rho_2\right)-S\left(\rho_{1,2}\right)=2 f\left(\sigma_1\right)-f\left(\sigma_{+}\right)-f\left(\sigma_{-}\right) ,
\end{equation}
where
\begin{equation}
f(x)=\left(\frac{x+1}{2}\right) \log _2\left(\frac{x+1}{2}\right)-\left(\frac{x-1}{2}\right) \log _2\left(\frac{x-1}{2}\right)
\end{equation}
for $x>1$.
    
    As expected, the mutual information decreases with the distance between two patches for all $\nu$ considered. Moreover,   as $\nu$ increases, the mutual information increases. Specifically, we notice higher enhancement of mutual information in constant rolls with large $\nu$ than the SR-USR pair, indicating that an insertion of such a constant roll would generally increase the probability of verifying quantumness of primordial perturbations.\footnote{The caveat for the signal enhancement in multi-phase de Sitter is that the 
    transition between different phases would significantly modify the entanglement signal \cite{Brahma:2024ycc}.}

    In principle, inflation with different kinds of phases will generate different quantum informational aspects, which may be detected by cosmic Bell tests \cite{Espinosa-Portales:2022yok}.

\subsection{Negativity}

Since we are examining the symplectic symmetry of different quasi-de Sitter backgrounds, we investigate another measure of the quantum correlation between two systems, namely the log negativity. 
Unlike measures like the mutual information, which is a function of the symplectic eigenvalues of the 
full covariance matrix $\gamma$, the log negativity is a function of the symplectic eigenvalues of the 
partially transposed matrix $\tilde{\gamma}$ which, in general, might be different for the Wands-duals even if $\gamma$
coincides for the two.
Let us define it explicitly as follows. 
Given a state $\rho_{AB}$ for the bipartite system, the partial transpose with respect to subsystem $B$ is given by
\begin{equation}
\rho_{AB}^{T_{B}}=\sum_{i,j,k,l}
\rho_{ij,kl}\,
|i\rangle_A \langle k|_A
\otimes
|l\rangle_B \langle j|_B
\end{equation}
where the composite density matrix reads
\begin{equation}
\rho_{AB}=\sum_{i,j,k,l}
\rho_{ij,kl}\,
|i\rangle_A \langle k|_A
\otimes
|j\rangle_B \langle l|_B\,,
\end{equation}
with $\{ \vert i\rangle_A \}$ and $\{ \vert j\rangle_B \}$ orthonormal bases of the respective subsystem 
Hilbert spaces. 
The log negativity is then defined as:
\begin{equation}
LN(\rho_{AB}) =
\log_2 \left[
\mathrm{Tr}\sqrt{
\left(\rho_{AB}^{T_B}\right)^\dagger
\rho_{AB}^{T_B}
}
\right]\,.
\end{equation}
The log negativity is invariant under local symplectic transformation. For Gaussian systems, $LN>0$ signals entanglement between the two subsystems, and admits a closed form expression in terms of the covariance matrix. For our case, the covariance matrix can be written as:
\begin{equation}
    \gamma =
\begin{pmatrix}
A & C\\
C^T & A
\end{pmatrix}\,,
\end{equation}
with the `entries' $A$ and $C$ obtained from the smeared two-point correlators, computed as momentum-space integrals of the spectra $P_{vv}$, $P_{v\pi}$, and $P_{\pi\pi}$.

For a two-mode Gaussian state, the logarithmic negativity is completely determined by the smallest symplectic eigenvalue $\tilde\sigma_-$ of the partially transposed covariance matrix $\tilde{\gamma}$:
\begin{equation}
    LN = \max\{0,-\ln \tilde\sigma_-\}.
\end{equation}
In turn, for the symmetric case $\tilde\sigma_-$ can be expressed in terms of the quantities
\begin{equation}
    \tilde\Delta = 2\det A - 2\det C,
\qquad
\det\tilde{\gamma}=\det\gamma,
\end{equation}
since
\begin{equation}
   \tilde\sigma_-=
\sqrt{
\frac{
\tilde\Delta-\sqrt{\tilde\Delta^2-4\det\gamma}
}{2}
}. 
\end{equation}

Numerically, we find $\tilde{\sigma}_{-} \gg 1$ for all Wands duals and across all separations for which the two patches are disjointed, so that $LN = 0$. This indicates that the two local Gaussian modes we have extracted are not entangled. 


The crucial observation for our purposes is that the logarithmic negativity is the same for any Wands dual pairs~\eqref{wands}, {\it i.e.}, $LN^{(I)} = LN^{(II)}$. Notably, this is not a corollary of the symplectic eigenvalue invariance established earlier in this work. Indeed, $LN$ is determined by the symplectic eigenvalues of the partially transposed matrix $\tilde\gamma$, which are in general not fixed by those of $\gamma$. Instead, the log nevativity equality across duals follows from their dependence on the determinants of the sub-blocks of the covariance matrix. As seen above, $\tilde\sigma_-$ depends on the covariance matrix only through the invariants $\det A$, $\det C$, and $\det\gamma$. Under a local canonical transformation $\gamma \longrightarrow \gamma^{\prime} = (S_A \oplus S_B) \gamma (S_A \oplus S_B)^T$, with $S_A, S_B \in \mathrm{Sp}(2,\mathbb{R})$, each of these determinants is preserved, since symplectic matrices have unit determinant, rendering the logarithmic negativity invariant. As elaborated in the Conclusions, the transformation relating two Wands dual backgrounds acts in precisely this form in our spatial entanglement setup.


	\section{Conclusions}
	\label{sec:conclusions}
	
	We have shown how inflationary backgrounds related by Wands duality share equivalent quantum-informational properties for localized observables.
    Using the continuous-variable formalism for coarse-grained scalar perturbations within the linear theory, we constructed the covariance matrix of a bipartite system formed by two spatially separated regions in quasi-de Sitter space and extracted its symplectic invariants. On reflection, this seems like a rather remarkable fact. Although the reduced power spectra and covariance matrices for Wands-dual	backgrounds can differ substantially, the symplectic eigenvalues of the bipartite covariance matrix are identical for the dual partners. We demonstrated this explicitly for the slow-roll/ultra-slow-roll pair and then proved it for a generic Wands dual pair.
    As a result, all quantum-informatic quantities associated with this bipartite Gaussian system  -- entanglement entropy, mutual information, quantum discord and negativity -- have been proved to be invariant under Wands duality. 
    
    The conceptual key to this result is the following observation. A Wands-duality transformation leaves the Mukhanov--Sasaki mode function $v_k$ invariant, and therefore also leaves $v_k'$ invariant. The map $(v_k, v_k') \to (v_k, \pi_k = v_k' - z'/z\, v_k)$ is a canonical transformation on phase space. From this perspective, one could have worked with $(v_k,v_k')$ as the canonical pair of variables from the outset: the Wands-dual backgrounds are then related by a linear, $k$-\emph{independent} canonical transformation that acts locally on each of the two coarse-grained subsystems. More explicitly, 
    a Wands dual transformation $M_{\mathrm{Wands}}:\left(v_k^{(I)}, \pi_k^{(I)}\right)\mapsto\left(v_k^{(II)}, \pi_k^{(II)}\right)$  is given by    \begin{equation}\label{eq:MWands}
M_{\mathrm{Wands}}=\left(\begin{array}{cc}
1 & 0 \\
-f^{(I)} & 1
\end{array}\right)\left(\begin{array}{cc}
1 & 0 \\
-f^{(II)} & 1
\end{array}\right)^{-1}
\end{equation}
with $f^{(I)}=\left(a^{\prime} / a\right)\left[1+(-3 + 2 \nu) / 2\right]$ and $f^{(II)}=\left(a^{\prime} / a\right)\left[1+(-3 - 2 \nu) / 2\right]$. Hence, the transformation between the local covariance matrices of a Wands dual pair, written as $T_{\mathrm{Wands}}: \gamma^{(I)}\mapsto\gamma^{(II)}$, is an element in $\mathrm{Sp}(2, \mathbb{R}) \times \mathrm{Sp}(2, \mathbb{R})$, rather than the full $\mathrm{Sp}(4,\mathbb{R})$ group. Since symplectic eigenvalues are invariant under any symplectic congruence $\gamma \to S\gamma S^T$, and since local symplectic transformations also commute with the partial transposition used to define logarithmic negativity, the entire Gaussian symplectic spectrum,  of both $\gamma$ and $\tilde\gamma$, is common to both members of a Wands-dual pair.

The $k$-independence of $M_{\rm Wands}$ is the crucial structural feature. Since the canonical transformation does not depend on the mode number $k$, it commutes with the momentum-space integral that builds the real-space covariance matrix, rendering the two dual backgrounds indistinguishable under all local Gaussian quantum-informational  probes. This also immediately implies that the result is independent of the choice of window function, since the window enters only through the above-mentioned integral. By contrast, a scale-dependent canonical transformation, would generically break this commutation and produce a real-space entanglement signature distinguishing it from the standard case. However, such transformations would not have an interpretation as any Wands' duality symmetry. The emerging lesson from all this is that Wands-dual backgrounds are physically equivalent at the Gaussian level because, and only because, the canonical transformation relating them is \emph{local} (in the bipartite sense) and \emph{scale-independent}. Any would-be entanglement measure capable of distinguishing Wands-dual backgrounds at Gaussian order would necessarily fail to be invariant under local, scale-independent canonical transformations, and would therefore not constitute a good measure of bipartite entanglement between the two patches.
In this sense, the quantum-informatic symmetry we identify is not an accidental feature of the specific backgrounds studied; it is a direct consequence of the group-theoretic structure of the Wands-duality transformation.

These findings extend the usual statement of Wands' duality -- that distinct inflationary histories share the same configuration-space power spectrum for the canonical Mukhanov-Sasaki variable -- to the full symplectic invariant content of the associated
Gaussian state. Our results also connect naturally with the geometric approach of \cite{Agullo:2024cln, Ribes-Metidieri:2025nfw}, where entanglement between local
observables is encoded in the eigenvalues of a positive-definite metric on phase space. In that language, the equality of symplectic spectra for Wands-dual backgrounds translate directly into equality of all invariants constructed from the metric/complex-structure pair. At the same time, it is important, however, to stress what this symmetry does \emph{not} imply. The physical curvature perturbation $\zeta$ is related to the Mukhanov-Sasaki variable by a
background-dependent canonical transformation, $v = z\zeta$ with $z(\eta) \propto a\sqrt{\epsilon_1}$ differing between Wands-dual backgrounds. Consequently, superhorizon $\zeta$ is conserved in slow-roll inflation but grows in ultra-slow-roll, which is consistent with our results. Our quantum-informatic symmetry is therefore a statement about the Gaussian state of the coarse-grained variables, and does not contradict the well-known inequivalence of the dynamics of the physical curvature perturbation between Wands dual backgrounds.

Several avenues for future work suggest themselves. On the formal side, the argument above shows that the result extends directly to multipartite Gaussian settings and is window-function independent, but it would be interesting to investigate further whether it
persists beyond the constant-$\nu$ approximation. An immediate open question is what happens when interactions or non-Gaussianities are included \cite{Micheli:2025yux}: Mode-coupling for non-Gaussian couplings would lead to additional complications when translating to real-space. Moreover, note that the SR-USR duality breaks down at non-linear order, as shown in \cite{Ireland:2026txt} using Wigner negativity, demonstrating that non-linear canonical transformations are no longer captured by elements of $\mathrm{Sp}(2N,\mathbb{R})$.
On the phenomenological side, it would be valuable to examine phase transitions between different quasi-de Sitter
backgrounds. This is a natural setting in which the scale independence underpinning the quantum-informatic invariance 
is implicitly broken. A transition at a conformal time $\eta_\star$ singles out the comoving scale $k_\star \sim a_\star H_\star$, with modes, inside, near and outside the horizon responding differently (in fact, non-adiabatically for $k\sim k_\star$). Thus, the effective canonical transformation relating pre- and post-transition Gaussian states should 
acquire an implicit $k$-dependence and could potentially lead to significant deviations from the Wands invariance
in the measures studied here, with overall different real-space entanglement signatures obtained for either SR or USR backgrounds. Along similar lines, the effect of excited initial states on our symmetry, and whether such states can enhance any of the entanglement measures, remains another important direction. Finally, it would be interesting to assess whether the quantum-informatic invariance under Wands duality has observable consequences for cosmic Bell tests extractable from the CMB \cite{Espinosa-Portales:2022yok}.


\section*{Acknowledgments}
S.B. is supported in part by a start-up grant from the Indian Statistical Institute. J.C.F. is funded by the STFC under grant number ST/X001040/1. X.L. is supported in part by the Program of China Scholarship Council (Grant No. 202208170014).


	\appendix
	
	
	\section{Window function and canonical rescaling}
	\label{app:window}
	
	For completeness we collect here the explicit expressions for the window function, its Fourier transform, and the function $G(\delta)$ entering the equal-point commutator, following Ref.~\cite{Martin:2021qkg}. We work with the spherically symmetric profile
	\begin{equation}
    \label{eq:window-function}
		W(x) = \frac{3}{4\pi F(\delta)}
		\begin{cases}
			1 & \text{for } x \le 1\,, \\
			- \dfrac{1}{\delta} (x-1) + 1 & \text{for } 1 < x \le 1+\delta , \\
			0 & \text{for } x > 1+\delta\,,
		\end{cases}
	\end{equation}
	where the normalization factor \(F(\delta)\) is given by
	\begin{equation}
		F(\delta) = \frac{1}{4(\delta+2)} \left( \delta^2 + 2\delta + 2 \right)\,.
	\end{equation}
	This ensures that
	\begin{equation}
		4\pi \int_0^{\infty} \mathrm{d}x \, x^2 W(x) = 1\,.
	\end{equation}
	
	The corresponding Fourier transform entering the coarse-grained modes is
	\begin{eqnarray}
		\widetilde{W}\!\left( \frac{kR}{a} \right)
		&=& \frac{3}{F(\delta)} \left( \frac{kR}{a} \right)^{-3}
		\left[
		\frac{1}{\delta} \sin\!\left( \frac{kR}{a} \right)
		- \left( 1 + \frac{1}{\delta} \right)
		\sin\!\left( (1+\delta) \frac{kR}{a} \right)\right.\nonumber\\
 & & \;\;\;\;\;\;\;\;\; \left.	+ \frac{2a}{\delta kR} \cos\!\left( \frac{kR}{a} \right)
		- \frac{2a}{\delta kR} \cos\!\left( (1+\delta) \frac{kR}{a} \right)
		\right]\,.
	\end{eqnarray}
	In the IR limit this reproduces the behaviour of a sharp top-hat, while in the UV	it is softened enough to avoid the divergences associated with a strict step function. The top-hat form is recovered in the limit $\delta \to 0$.
	
	Using this window, the equal-point commutator of coarse-grained fields reads
	\begin{equation}
		\big[ \hat{q}_{R,1}(\mathbf{x}), \hat{q}_{R,2}(\mathbf{x}) \big]
		= i 4\pi \left( \frac{a}{R} \right)^3
		\int_0^{\infty} \mathrm{d}u \, u^2 W^2(u)
		= i \, \frac{3}{4\pi} \left( \frac{a}{R} \right)^3 G(\delta)\,,
	\end{equation}
	where
	\begin{equation}
		G(\delta)
		= \frac{8}{5(\delta+2)^2 (\delta^2 + 2\delta + 2)^2}
		\left( \delta^3 + 5\delta^2 + 10\delta + 10 \right)\,.
	\end{equation}
	This is the function that enters the rescaling matrix $\Lambda^{(1)}$.


		\section{Explicit symplectic eigenvalues for a given CR pair}		\label{app:explicit_sigmas}

	For the CR Wands-dual pair with \(\nu=5/2\), it is convenient to introduce a generalized	set of window integrals adapted to the higher powers of \((k\eta)^{-1}\) appearing in the power spectra:
	\begin{equation}
		\mathcal{K}^{(\pm)}_{\mu,n}
		\equiv \int_{\beta}^{\infty}\! \mathrm{d}w\, w^{\mu-2n}\,\widetilde{W}^2(w)
		\;\pm\;
		\int_{\beta}^{\infty}\! \mathrm{d}w\, w^{\mu-2n}\,\widetilde{W}^2(w)\,\mathrm{sinc}(\alpha w),
		\qquad \mu\in\{1,3\},\quad n=0,1,2,3.
	\end{equation}
	Inserting the CR power spectra into ~\eqref{eq:gamma_general}, and organizing the resulting $\gamma^{\mathrm{I,II}}_{ab}$ in terms of these integrals, one finds that the squared symplectic eigenvalues in both CR realizations can be written as
	\begin{equation}
		\left(\sigma_\pm^{\mathrm{I}}\right)^2
		= \left(\sigma_\pm^{\mathrm{II}}\right)^2
		= \frac{4(HR)^2}{9\pi^2 G(\delta)^2}\,
		\mathcal{F}_\pm^{(\nu=5/2)}\big(\{\mathcal{K}^{(\pm)}_{\mu,n}\}\big)\,,
	\end{equation}
	where $\mathcal{F}_\pm^{(\nu=5/2)}$ is an explicit polynomial in the symbols $\mathcal{K}^{(\pm)}_{\mu,n}$, obtained by inserting the matrix elements \(\gamma_{ab}^{\mathrm{I,II}}\) into ~\eqref{eq:sigma_pm}. Although the full expression is algebraically lengthy, it is unique and common to both members of	the CR dual pair, such that
	\begin{equation}
		\left(\sigma_\pm^{\mathrm{I}}\right)^2
		= \left(\sigma_\pm^{\mathrm{II}}\right)^2
		\equiv \left(\sigma_\pm^{(\nu=5/2)}\right)^2\,.
	\end{equation}
	This establishes explicitly that the CR Wands-dual backgrounds also share identical	symplectic spectra, in agreement with the general argument of Sec.~\ref{subsec:general_duals}.
	
	
	\section{General Wands duals and Hankel integrals}
	\label{app:general_duals}
	
	In this appendix, we spell out the Hankel-function manipulations underlying the invariance of the quantum-informatic measures for general Wands-dual pairs, as argued in Sec.~\ref{subsec:general_duals}.	We consider the quasi-de Sitter mode functions
	\begin{equation}
		v_k(\eta) = \frac{1}{2}
		e^{i\frac{\pi}{2}(\nu+1/2)}\sqrt{-\pi\eta}\,
		H^{(1)}_{\nu}(-k\eta)\,,
	\end{equation}
	and the conjugate momenta for the two Wands-dual realizations
	\begin{align}
		\pi_k^{(I)}(\eta) &= -\frac{1}{2}\sqrt{k\pi}\,
		e^{i\frac{\pi}{2}(\nu+1/2)}(-k\eta)H^{(1)}_{\nu-1}(-k\eta)\,, \\
		\pi_k^{(II)}(\eta) &= \frac{1}{2}\sqrt{k\pi}\,
		e^{i\frac{\pi}{2}(\nu+1/2)}(-k\eta)H^{(1)}_{\nu+1}(-k\eta)\,,
	\end{align}
	corresponding to $\epsilon_2=-3\pm 2\nu$.
	
	From these expressions, the reduced power spectra for the $+$ and $-$ realizations can be computed explicitly. For the $+$ case, one finds
	\begin{align}
		P^{(I)}_{vv}(k) &= -\frac{k^3\eta}{8\pi}
		H^{(1)\,*}_{\nu}(-k\eta)\,H^{(1)}_{\nu}(-k\eta)\,, \\
		P^{(I)}_{v\pi}(k) &= \frac{k^4\eta}{16\pi}
		\Big[ H^{(1)\,*}_{\nu}(-k\eta)\,H^{(1)}_{\nu-1}(-k\eta)
		+ H^{(1)\,*}_{\nu-1}(-k\eta)\,H^{(1)}_{\nu}(-k\eta) \Big]\,, \\
		P^{(I)}_{\pi\pi}(k) &= -\frac{k^5\eta}{8\pi}
		H^{(1)\,*}_{\nu-1}(-k\eta)\,H^{(1)}_{\nu-1}(-k\eta)\,,
	\end{align}
	while for the $-$ case one obtains
	\begin{align}
		P^{(II)}_{vv}(k) &= -\frac{k^3\eta}{8\pi}
		H^{(1)\,*}_{\nu}(-k\eta)\,H^{(1)}_{\nu}(-k\eta)\,, \\
		P^{(II)}_{v\pi}(k) &= -\frac{k^4\eta}{16\pi}
		\Big[ H^{(1)\,*}_{\nu+1}(-k\eta)\,H^{(1)}_{\nu}(-k\eta)
		+ H^{(1)\,*}_{\nu}(-k\eta)\,H^{(1)}_{\nu+1}(-k\eta) \Big]\,, \\
		P^{(II)}_{\pi\pi}(k) &= -\frac{k^5\eta}{8\pi}
		H^{(1)\,*}_{\nu+1}(-k\eta)\,H^{(1)}_{\nu+1}(-k\eta)\,.
	\end{align}
	In both cases the $vv$-configuration field spectrum is the same, as expected from the fact that the two realizations share the same Hankel index $\nu$.
	
	To connect the two sets of spectra, we use the Hankel recurrence relation
	\begin{equation}
		H^{(1)}_{\nu}(-k\eta) = \frac{-k\eta}{2\nu}\left[ H^{(1)}_{\nu-1}(-k\eta) + H^{(1)}_{\nu+1}(-k\eta) \right]\,.
	\end{equation}
	This allows all occurrences of $H^{(1)}_{\nu}$ to be eliminated in favour of \(H^{(1)}_{\nu\pm 1}\), so that both $P^{(I)}_{ij}$ and $P^{(II)}_{ij}$ can be rewritten as linear combinations of products of the	form $H^{(1)}_{\nu-1}H^{(1)\,*}_{\nu-1}$, $H^{(1)}_{\nu+1}H^{(1)\,*}_{\nu-1}$, $H^{(1)}_{\nu-1}H^{(1)\,*}_{\nu+1}$, and $H^{(1)}_{\nu+1}H^{(1)\,*}_{\nu+1}$.
	
	When these spectra are inserted into the expression of the covariance matrix ~\eqref{eq:gamma_general}, the dependence on the Hankel functions enters only through four basic integrals of the type
	\begin{align}
		A &= \int \mathrm{d}k\,f(k)\,H^{(1)}_{\nu-1}(-k\eta)\,H^{(1)\,*}_{\nu-1}(-k\eta)\,, \\
		B &= \int \mathrm{d}k\,f(k)\,H^{(1)}_{\nu+1}(-k\eta)\,H^{(1)\,*}_{\nu-1}(-k\eta)\,, \\
		C &= \int \mathrm{d}k\,f(k)\,H^{(1)}_{\nu-1}(-k\eta)\,H^{(1)\,*}_{\nu+1}(-k\eta)\,, \\
		D &= \int \mathrm{d}k\,f(k)\,H^{(1)}_{\nu+1}(-k\eta)\,H^{(1)\,*}_{\nu+1}(-k\eta)\,,
	\end{align}
	where $f(k)$ collects all remaining factors (window function, normalization factor, numerical coefficients,and so on) which appeared in calculating the ${\bf \gamma}$.	The different components of the two covariance matrices ${\bf \gamma}^{(I)}$ and ${\bf \gamma}^{(II)}$ can then be expressed in terms of $A,B,C,D$.
	
	Choosing $\lambda = \sqrt{a/R}$ for definiteness, one finds schematically that 
	\begin{align}
		\gamma^{(I)}_{11} &= \gamma^{(II)}_{11}
		= -\frac{\eta^3}{32\pi\nu^2}(A+B+C+D)\,, \\
		\gamma^{(I)}_{12} &= -\frac{\eta^2}{32\pi\nu}(2A+B+C)\,, \\
		\gamma^{(II)}_{12} &= \frac{\eta^2}{32\pi\nu}(B+C+2D)\,, \\
		\gamma^{(I)}_{22} &= -\frac{\eta}{8\pi}A\,, \\
		\gamma^{(II)}_{22} &= -\frac{\eta}{8\pi}D\,,
	\end{align}
	with the remaining entries can be obtained in exactly the same way but with $f(k)$ replaced by $g(k) = f(k)\,\mathrm{sinc}(kd/a)$ due to the off-diagonal structure in ~\eqref{eq:gamma_general}.
	
	Inserting these expressions into the symplectic invariants defined in	Eqs.~\eqref{eq:sigma_pm}-\eqref{eq:sigma12}, one finds that the combinations appearing in $\sigma_\pm^2$, $\sigma_1^2$ and $\sigma_{1-2}^2$ are the same for the $I$ and $II$	realizations.
	In other words,
	\begin{equation}
		\sigma^{(I)}_{\pm} = \sigma^{(II)}_{\pm}\,,\qquad
		\sigma^{(I)}_1 = \sigma^{(II)}_1\,,\qquad
		\sigma^{(I)}_{1-2} = \sigma^{(II)}_{1-2}\,.
	\end{equation}
	This establishes that, for any pair of Wands-dual backgrounds sharing the same Hankel index $\nu$, all symplectic invariants of the coarse-grained bipartite system coincide, in agreement with the general statement in Sec.~\ref{subsec:general_duals}.
	

\printbibliography

\end{document}